\documentclass[11pt,a4paper]{article}

\usepackage{mathrsfs}
\usepackage{amsmath,amssymb}
\usepackage{hyperref}
\usepackage{verbatim}
\usepackage{stmaryrd}
\usepackage{amsfonts}

\def\S{{\mathbb S}}
\def\K{{\mathbb K}}
\def\R{{\mathbb R}}
\def\L{{\mathbb L}}
\def\Q{{\mathbb Q}}
\def\G{{\mathbb G}}
\def\H{{\mathbb H}}
\def\B{{\mathbb U}}
\def\C{{\mathbb C}}

\def\Uq{U_{q}(\mathfrak{g})}

\newcommand{\alg}[1]{\mathfrak{#1}}

\newcommand{\el}{\nonumber\\}

\def\ads{{\rm AdS}_5\times {\rm S}^5}

\def\h{\mathfrak{a}}
\def\g{\mathfrak{g}}
\def\m{\mathfrak{b}}

\def\afQ{\widehat{\cal{Q}}}
\def\afB{\widehat{\cal{B}}}

\def\ads{{\rm AdS}_5\times {\rm S}^5}

\usepackage[a4paper,text={450pt,650pt},centering]{geometry}

\begin{document}


\begin{titlepage}


\begin{centering}

\vspace{1in}

{\Large {\bf Coideal Quantum Affine Algebra and Boundary Scattering \\of the Deformed Hubbard Chain}}

\vspace{.5in}

{\large Marius de Leeuw${}^{a,b,1}$, Takuya Matsumoto${}^{c,d,2}$ and Vidas Regelskis${}^{e,f,3}$}\\
\vspace{.3 in}
${}^{a}${\emph{ETH Z\"urich, Institut f\"ur Theoretische Physik, \\Wolfgang-Pauli-Str.\ 27, CH-8093 Zurich, Switzerland}} \\
${}^{b}${\emph{Max-Planck-Institut f\"ur Gravitationsphysik\\Albert-Einstein-Institut\\Am M\"uhlenberg 1, 14476 Potsdam, Germany}} \\
${}^{c}${\emph{School of Mathematics and Statistics, University of Sydney, \\ NSW 2006, Australia}}\\
${}^{d}${\emph{Graduate School of Mathematics, Nagoya University,\\Nagoya 464-8602, Japan}}\\
${}^{e}${\emph{Department of Mathematics, University of York,\\Heslington, York YO10 5DD, UK}}\\
${}^{f}${\emph{Institute of Theoretical Physics and Astronomy of Vilnius University,\\Go\v{s}tauto 12, Vilnius 01108, Lithuania}}\\
%

\footnotetext[1]{{\tt deleeuwm@phys.ethz.ch,}\quad ${}^{2}${\tt tmatsumoto@usyd.edu.au}\quad ${}^{3}${\tt vr509@york.ac.uk}}
\vspace{.5in}

{\bf Abstract}

\vspace{.1in}
\end{centering}

{We consider boundary scattering for a semi-infinite one-dimensional deformed Hubbard chain with boundary conditions of the same type as for the Y=0 giant graviton in the AdS/CFT correspondence. We show that the recently constructed quantum affine algebra of the deformed Hubbard chain has a coideal subalgebra which is consistent with the reflection (boundary Yang-Baxter) equation. We derive the corresponding reflection matrix and furthermore show that the aforementioned algebra in the rational limit specializes to the (generalized) twisted Yangian of the Y=0 giant graviton.}

\end{titlepage}

\tableofcontents


\section{Introduction}

The Hubbard model was introduced in order to study strongly correlated electrons \cite{Hubbard}, and later due to many generalizations (e.g. \cite{HubSUN,AB,DFFR}) it grew into a large family of models (see e.g. \cite{Hub1,Hub2}). Recently, the interest in the Hubbard model has been renewed due to remarkable successes in solving similar models appearing in the context of the $\ads$ correspondence (see for instance review \cite{AdSReview} and references therein). 

An interesting relation between the Hubbard model and the $\ads$ string was found by studying the centrally extended $U_q(\mathfrak{su}(2|2))$ algebra $\mathcal{Q}$ \cite{BK}. The integrable model with this underlying algebra turns out to describe a variety of quantum deformed Hubbard models as well as the $\ads$ superstring in the rational $q\to1$ limit. This makes it an interesting model to study since it offers a unified description of all of these systems. We will simply refer to it as the deformed Hubbard model. 

It was found that the deformed Hubbard model is actually invariant under the affine extension, $\afQ$, of the symmetry algebra \cite{BGM}. In the rational $q\to1$ limit this algebra becomes the Yangian symmetry of the $\ads$ superstring \cite{BeisertYangian}. The fundamental R-matrix can be found by requiring invariance under $\mathcal{Q}$ alone, but the affine extension plays a crucial role in the determination of S-matrices in higher representations. The fundamental S-matrix was found in \cite{BK} and in the rational limit is equivalent to Shastry's R-matrix \cite{Shastry,BAnalytic,MartinsMelo}. The S-matrix describing bound state scattering has also recently been derived \cite{LMR}. Its construction relies heavily on the affine generators in $\afQ$. This is similar to the situation in $\ads$, where this part was played by the Yangian charges instead \cite{ALT}. 

When studying integrable models with periodic boundary conditions, the spectrum is governed by the S-matrix and thus indirectly through the underlying (bulk) symmetry algebra. However, for integrable systems with boundaries, there is another object, called the reflection matrix, which describes the scattering of excitations off the boundary. Generically, boundaries preserve a subalgebra of the bulk Lie algebra and this subalgebra then determines the corresponding reflection matrix. However this is usually not enough to determine the bound state reflection matrix and a coideal subalgebra of the corresponding bulk Yangian or quantum affine algebra is required.

Open boundary conditions for the one-dimensional Hubbard model have received less attention than their closed chain counterpart, but they exhibit a rich variety of structures (see e.g.\ \cite{HubOp1,HubOp2,HubOp3}). The reflection matrices for open boundary conditions for the deformed Hubbard chain have been studied in \cite{MN}. Drawing from similarities with open spin chains in the context of AdS/CFT \cite{HM}, two types of boundary conditions were formulated corresponding to $Y=0$ and $Z=0$ brane configurations. In this paper we will study boundary conditions corresponding to the $Y=0$ system applied to bound states. For this configuration, the boundary representation is a singlet and the boundary conditions preserve half of the supersymmetries.

The aim of this paper is twofold. Firstly, we want to identify the symmetry algebra that governs boundary scattering. For the $\ads$ superstring this algebra is a (generalized) twisted Yangian \cite{AN,MR1}. Here we find it to be an affine coideal subalgebra $\afB$ of $\afQ$. Its structure turns out to be governed by the notion of quantum symmetric pairs which have been heavily studied in the context of semisimple Lie algebras \cite{L1,L2,L3,L4}. Inspired by these results we explicitly construct the boundary algebra $\afB$. 

Secondly, having found the symmetry algebra $\afB$ we use it to compute the reflection matrix of arbitrary bound states and show that it satisfies the reflection (also called boundary Yang-Baxter) equation. Conversely, we explicitly solve the reflection equation and find that our reflection matrix is the unique solution, thus proving that $\afB$ is indeed the correct and unique symmetry algebra. Finally, we also show that in the $q\to1$ limit we reproduce the results for the $Y=0$ configuration for the $\ads$ superstring found in \cite{LP}.

It is worth to note, that somewhat similar boundary scattering problems for quantum affine algebras of the Lie algebras of classical type have been considered in \cite{MR,MRS}, where again the scattering is governed by some coideal subalgebra $\mathcal{B}$, which in some cases is called a $q$-Yangian \cite{LM}. Boundary scattering has also been intensively studied for sine-Gordon and affine Toda field theories \cite{MeN,DM,DG,BB}. The investigation of reflection equation bearing on the quantum symmetric pairs constructed by G. Letzter was considered in \cite{Kolb,KS}. 

This work is organized as follows. In section 2 we discuss the relevant notation and definitions of the bulk algebra and its bound state representations. In section 3 we present the required axiomatic formulation of the coideal subalgebras and quantum symmetric pairs. Then in section 4 we present the general form of reflection matrix for arbitrary bound states and discuss its properties. We end with some concluding remarks. The Appendix is reserved for the $q\to1$ limit of $\afQ$ and also for a brief review of the twisted Yangian of the $Y=0$ giant graviton.


\section{Deformed quantum affine algebra}\label{sec:Q}

In this section we shall review the quantum affine algebra constructed
in \cite{BGM} and its bound state representation constructed in \cite{LMR}.


\paragraph{Quantum affine algebra $\afQ$.}

The algebra $\widehat{\mathcal{Q}}$ of the quantum deformed one-dimensional Hubbard
chain was recently constructed in \cite{BGM} and is a deformation of the centrally extended affine algebra $\widehat{\mathfrak{sl}}(2|2)$.
It is generated by four sets of the Chevalley-Serre generators $K_i\equiv q^{H_i}$,
$E_i$, $F_i$ ($i=1,\,2,\,3,\,4$) and two sets of the central elements
$U_{k}$ and $V_{k}$ ($k=2,\,4$) with $U_{k}$ being responsible for the deformation of the coproduct. 

Let us start by recalling the symmetric
matrix $DA$ and the normalization matrix $D$ associated to the Cartan
matrix $A$ for $\widehat{\mathfrak{sl}}(2|2)$:
\begin{equation}
DA=\begin{pmatrix}2 & -1 & 0 & -1\\
-1 & 0 & 1 & 0\\
0 & 1 & -2 & 1\\
-1 & 0 & 1 & 0
\end{pmatrix},\qquad D=\mathrm{diag}(1,-1,-1,-1).\label{DA}
\end{equation}
The algebra is then defined accordingly by the following commutation relations,
\begin{align}
 & K_{i}E_{j}=q^{DA_{ij}}E_{j}K_{i}, &  & K_{i}F_{j}=q^{-DA_{ij}}F_{j}K_{i},\el
 & \{E_{2},F_{4}\}=-\tilde{g}\tilde{\alpha}^{-1}(K_{4}-U_{2}U_{4}^{-1}K_{2}^{-1}), &  & \{E_{4},F_{2}\}=\tilde{g}\tilde{\alpha}^{+1}(K_{2}-U_{4}U_{2}^{-1}K_{4}^{-1}),\el
 & [E_{j},F_{j}\}=D_{jj}\frac{K_{j}-K_{j}^{-1}}{q-q^{-1}}, &  & [E_{i},F_{j}\}=0,\quad i\neq j,\ i+j\neq6.
\end{align}
These are supplemented by a set of Serre relations ($j=1,\,3$) 
\begin{align}
 & [E_{j},[E_{j},E_{k}]]-(q-2+q^{-1})E_{j}E_{k}E_{j}=0, && [E_{1},E_{3}]=E_{2}E_{2}=E_{4}E_{4}=\{E_{2},E_{4}\}=0, \el
 & [F_{j},[F_{j},F_{k}]]-(q-2+q^{-1})F_{j}F_{k}F_{j}=0, && [F_{1},F_{3}]=F_{2}F_{2}=F_{4}F_{4}=\{F_{2},F_{4}\}=0.
\end{align}
Central elements are related to the quartic Serre relations (for $k=2,\,4$) as follows,
\begin{align}
 & \{[E_{1},E_{k}],[E_{3},E_{k}]\}-(q-2+q^{-1})E_{k}E_{1}E_{3}E_{k}=g_{k}\alpha_{k}(1-V_{k}^{2}U_{k}^{2}),\el
 & \{[F_{1},F_{k}],[F_{3},F_{k}]\}-(q-2+q^{-1})F_{k}F_{1}F_{3}F_{k}=g_{k}\alpha_{k}^{-1}(V_{k}^{-2}-U_{k}^{-2}).
\end{align}
This algebra has three central charges,
\begin{eqnarray}
C_{1} & = & K_{1}K_{2}^{2}K_{3},\el
C_{2} & = & \{[E_{2},E_{1}],[E_{2},E_{3}]\}-(q-2+q^{-1})E_{2}E_{1}E_{3}E_{2},\el
C_{3} & = & \{[F_{2},F_{1}],[F_{2},F_{3}]\}-(q-2+q^{-1})F_{2}F_{1}F_{3}F_{2}.\label{C123}
\end{eqnarray}
The central elements $V_{k}$ are constrained by the relation $K_{1}^{-1}K_{k}^{-2}K_{3}^{-1}=V_{k}^{2}.$ 


\paragraph{Hopf algebra.}

The group-like elements $X\in\{1,K_{j},U_{k},V_{k}\}$ ($j=1,2,3,4$ and $k=2,4$) have the coproduct $\Delta$ defined in a usual way, $\Delta(X)=X\otimes X$, while for the remaining Chevalley-Serre generators they are deformed by the central elements $U_{k}$. Similar considerations work for the antipode ${\rm S}$ and co-unit $\varepsilon$. Summarizing we have
\begin{equation}
\Delta(E_{j})=E_{j}\otimes1+K_{j}^{-1}U_{2}^{+\delta_{j,2}}U_{4}^{+\delta_{j,4}}\otimes E_{j},\quad\Delta(F_{j})=F_{j}\otimes K_{j}+U_{2}^{-\delta_{j,2}}U_{4}^{-\delta_{j,4}}\otimes F_{j}.\label{copEF}
\end{equation}


\paragraph{Representation.}

We shall be using the $q$-oscillator representation (for any complex $q$ not a root of unity) constructed in
\cite{LMR}. The bound state representation is defined on vectors 
\begin{align}
|m,n,k,l\rangle=(\mathsf{a}_{3}^{\dag})^{m}(\mathsf{a}_{4}^{\dag})^{n}(\mathsf{a}_{1}^{\dag})^{k}(\mathsf{a}_{2}^{\dag})^{l}\,|0\rangle,
\end{align}
where the indices $1$, $2$ denote bosonic and $3$, $4$ - fermionic oscillators;
the total number of excitations $k+l+m+n=M$ is the bound state
number and the dimension of the representation is dim$\,=4M$. 
This representation constrains the central elements as $U:=U_2=U^{-1}_4$ 
and $V:=V_2=V^{-1}_4$ and describes a spin-chain excitation with 
quasi-momentum $p$ related to the deformation parameter as $U=e^{ip}$. 

The triples corresponding to the bosonic and fermionic $\mathfrak{sl}_{q}(2)$
in this representation are given by 
\begin{align}
 & H_{1}|m,n,k,l\rangle=(l-k)|m,n,k,l\rangle, &  & H_{3}|m,n,k,l\rangle=(n-m)|m,n,k,l\rangle,\el
 & E_{1}|m,n,k,l\rangle=[k]_{q}\,|m,n,k-1,l+1\rangle, &  & E_{3}|m,n,k,l\rangle=|m+1,n-1,k,l\rangle,\el
 & F_{1}|m,n,k,l\rangle=[l]_{q}\,|m,n,k+1,l-1\rangle, &  & F_{3}|m,n,k,l\rangle=|m-1,n+1,k,l\rangle.
\end{align}
The supercharges act on basis states as 
\begin{align}
H_{2}|m,n,k,l\rangle= & ~-\left\{ C-\frac{k-l+m-n}{2}\right\} |m,n,k,l\rangle,\el
E_{2}|m,n,k,l\rangle= & ~a~(-1)^{m}[l]_{q}\,|m,n+1,k,l-1\rangle+b~|m-1,n,k+1,l\rangle,\el
F_{2}|m,n,k,l\rangle= & ~c~[k]_{q}\,|m+1,n,k-1,l\rangle+d~(-1)^{m}\,|m,n-1,k,l+1\rangle.
\end{align}
Here $[n]_q=(q^n-q^{-n})/(q-q^{-1})$ and $C$ is the $q$-factor of the central element $V = q^C$ and represents the energy of the state. The representation labels $a,b,c,d$ satisfy constraints 
\begin{align}
 & ad=\frac{q^{\frac{M}{2}}V-q^{-\frac{M}{2}}V^{-1}}{q^{M}-q^{-M}}\,, && bc=\frac{q^{-\frac{M}{2}}V-q^{\frac{M}{2}}V^{-1}}{q^{M}-q^{-M}}\,, \el
 & ab=\frac{g\alpha}{[M]_{q}}(1-U^{2}V^{2})\,, && cd=\frac{g\alpha^{-1}}{[M]_{q}}(V^{-2}-U^{-2})\,, \label{rep}
\end{align}
which altogether give the multiplet shortening (mass--shell) condition 
\begin{equation} \label{short}
\frac{g^{2}}{[M]_{q}^{2}}(V^{-2}-U^{-2})(1-U^{2}V^{2})=\frac{(V-q^{M}V^{-1})(V-q^{-M}V^{-1})}{(q^{M}-q^{-M})^{2}}\;.
\end{equation}
The explicit $x^{\pm}$ parametrization of the representation labels is 
\begin{align}
a & =\sqrt{\frac{g}{[M]_{q}}}\gamma\,, && b=\sqrt{\frac{g}{[M]_{q}}}\frac{\alpha}{\gamma}\frac{x^{-}-x^{+}}{x^{-}}\,,\el
c & =\sqrt{\frac{g}{[M]_{q}}}\frac{\gamma}{\alpha\, V}\frac{i\,\tilde{g}\, q^{\frac{M}{2}}}{g(x^{+}+\xi)}\,, &  & d=\sqrt{\frac{g}{[M]_{q}}}\frac{\tilde{g}\, q^{\frac{M}{2}}V}{i\, g\,\gamma}\frac{x^{+}-x^{-}}{\xi x^{+}+1}\,.\label{abcd}
\end{align}
The central elements in this parametrization read as
\begin{align}
&U^2 = \frac{1}{q^M} \frac{x^+ + \xi}{x^- + \xi} =  q^M \frac{x^+}{x^-}\frac{\xi x^- + 1}{\xi x^+ + 1}, 
  && V^2 = \frac{1}{q^M} \frac{\xi x^+ + 1}{\xi x^- + 1} = q^M \frac{x^+}{x^-}\frac{x^- + \xi}{x^+ + \xi},
\end{align}
while the shortening condition \eqref{short} becomes
\begin{align}
 \frac{1}{q^{M}}\left(x^+ + \frac{1}{x^+}\right)-q^M\left(x^- + \frac{1}{x^-}\right) = \left(q^M-\frac{1}{q^M}\right) \left(\xi+\frac{1}{\xi}\right),
\end{align}
here $\xi = -i \tilde g (q-q^{-1})$ and $\tilde g^2={g^2}/({1-g^2(q-q^{-1})^2})$.

The action of the affine charges $H_{4}$, $E_{4}$, $F_{4}$ is defined
in exactly the same way as for the regular supercharges subject to the following
substitutions $C\to-C$ and $(a,b,c,d)\to(\tilde{a},\tilde{b},\tilde{c},\tilde{d}).$
Then the affine labels $\tilde{a},\tilde{b},\tilde{c},\tilde{d}$ are acquired
from (\ref{abcd}) by replacing
%
\begin{align}
V\rightarrow V^{-1}, \qquad 
x^{\pm}\rightarrow\frac{1}{x^{\pm}}, \qquad 
\gamma\rightarrow\frac{i\tilde{\alpha}\gamma}{x^{+}}, \qquad 
\alpha\rightarrow\alpha\,\tilde{\alpha}^{2},  \qquad 
\tilde \alpha\rightarrow -\frac{1}{\tilde \alpha}\;.
\end{align}
Finally, we introduce the multiplicative spectral parameter of the algebra
\begin{align}\label{z}
z = \frac{1-U^2V^2}{V^2-U^2}=q^{-M}\vartheta(x^+)=q^{+M}\vartheta(x^-)
\quad\text{with}\quad 
\vartheta(x)=-\frac{(x + \xi) (1+1/(\xi x) )}{\xi-\xi^{-1}},
\end{align}
which will play an important role in describing the reflection algebra.


\section{Coideal quantum affine algebra}

We define the boundary conditions for the deformed one-dimensional Hubbard chain to be of the same type as those of the $Y=0$ giant graviton \cite{MN,HM} (see Appendix \ref{appYD3} for details). This kind of boundary conditions are described by a quantum version of the symmetric pair $(\mathfrak{g},\mathfrak{g}^{\theta})$ and the boundary scattering is governed by a coideal subalgebra. The axiomatic formulation and classification of coideal subalgebras and quantum symmetric pairs for semisimple Lie algebras was done by G. Letzter in series of works \cite{L1,L2,L3,L4}. We shall explicitly construct the quantum affine coideal subalgebra $\widehat{\mathcal{B}}$ of $\widehat{\mathcal{Q}}$ relying on G. Letzter's construction.


\subsection{Boundary algebra and symmetric pairs}

Before moving on to quantum deformed (affine) algebras, let us first briefly recall the algebraic structure for boundaries with Yangian algebras. Consider an integrable model with symmetry algebra described by the Yangian $\mbox{Y}(\g)$ for some Lie algebra $\g$. Suppose that the boundary module respects a subalgebra $\h\subset\g$ and that there is an involution $\theta:\g\mapsto\g$ such that $\h=\g^\theta$ is a $\theta$-fixed subalgebra of $\g$. Then $\h$ and subset $\m=\g \backslash \h$ respecting
\begin{align}\label{sym}
&[\h,\h] \subset \h, &&[\h,\m] \subset \m, &&[\m,\m] \subset \h.
\end{align}
are positive and negative eigenspaces of $\theta$, namely $\theta(\h)=+\h$ and $\theta(\m)=-\m$.
Thus, if the underlying symmetry algebra in the bulk is the Yangian $\mbox{Y}(\g)$, then the associated symmetry algebra respected by the boundary is the so-called (generalized) twisted Yangian $Y(\g,\h)$ \cite{DMS} generated by the level-0 charges $\mathbb{J}^{i}$ and twisted level-1
charges
\footnote{This construction is not valid when $\theta$ is trivial, $\g^\theta=\g$. For this case we refer to \cite{MR2}.}
\begin{equation}
\widetilde{\mathbb{J}}^{p}
:=\widehat{\mathbb{J}}^{p}+\frac{\alpha}{4}f_{\; qi}^{p}\,(\mathbb{J}^{q}\,\mathbb{J}^{i}+\mathbb{J}^{i}\,\mathbb{J}^{q})
=\widehat{\mathbb{J}}^{p} - \frac{\alpha}{4}[\,\mathbb{T}^{\alg{a}},\mathbb{J}^p],
\label{twist}
\end{equation}
where indices $i(,j,k,...)$ run over the $\h$-indices and $p,q(,r,...)$
over the $\m$-indices, $\alpha$ is a formal deformation
parameter conventionally set to $1$, and $\mathbb{T}^{\alg{a}}$ is the quadratic Casimir operator of $\alg{g}$
restricted to the subalgebra $\alg{a}$. 
The Yangian $\mbox{Y}(\g,\h)$ is a left coideal subalgebra,
\begin{align}
\Delta \mbox{Y}(\g,\h) \subset \mbox{Y}(\g) \otimes \mbox{Y}(\g,\h).
\end{align}
and is invariant under the extension $\bar{\theta}$ of the involution
$\theta$ acting on Yangian charges as 
\begin{equation}
\bar{\theta}(\mathbb{J}_{n}^{i})=(-1)^{n}\mathbb{J}_{n}^{i},\quad\bar{\theta}(\mathbb{J}_{n}^{p})=(-1)^{n+1}\mathbb{J}_{n}^{p},\quad\bar{\theta}(\alpha)=-\alpha,\label{sb}
\end{equation}
where $n$ is the level of the charge, thus it is easy to see that
$\tilde{\mathbb{J}}^{p}$ is invariant under $\bar{\theta}$ which acts as a filtration on $\mbox{Y}(\g)$.

However, this construction does not straightforwardly extend to quantum deformed algebras. Of course, at the level of the algebra, one can again define the involution $\theta$ that will specify the preserved subalgebra. But this cannot be extended to $\Uq$, since $U_q(\g^\theta)$ in the general case need not be a Hopf subalgebra of $\Uq$. This complicates identifying the symmetry algebra of the boundary which should clearly be a subalgebra of the full symmetry algebra.


\subsection{Coideal subalgebras and quantum symmetric pairs \label{sub:Coideals}}

We continue by giving the formulation of coideal subalgebras for semisimple algebras as described in \cite{L1,L2,L3,L4}. The results presented here allow for a generalization consistent with the affine structure presented in Section \ref{sec:Q}. We shall follow \cite{L3} quite closely and for the reader's convenience we shall try to give all the necessary constructions that will be used in later on in the explicit construction of the coideal subalgebra of $\afQ$. 


\paragraph{Setting.}

Consider a Lie algebra $\mathfrak{g}$ with Cartan matrix $(a_{ij})$
and a diagonal normalization matrix $(d_{i})$ (here and further $1\leq i,j\leq n$)
such that $(d_{i}a_{ij})$ is a symmetric matrix. Let the triangular
decomposition of the algebra be $\mathfrak{n}^{-}\oplus\mathfrak{h}\oplus\mathfrak{n}^{+}$
and let $\Phi$ denote the root system of $\mathfrak{g}$ and $\Phi^{+}$
 the set of the positive roots. Let $\pi=\{\alpha_{1},\alpha_{2},\ldots,\alpha_{n}\}$
be the set of simple positive roots and $(\alpha_{i},\alpha_{j})=d_{i}a_{ij}$ 
denote the Cartan inner product on $\mathfrak{h}^{*}$.
Then $\mathfrak{n}^+$ and $\mathfrak{n}^-$ have a basis of root
vectors $\{e_{\beta}\,|\,\beta\in\Phi^{+}\}$ and $\{f_{-\beta}\,|\,\beta\in\Phi^{+}\}$
respectively. Let $h_{1},\ldots,h_{n}$ be the basis of $\mathfrak{h}$.
Then the standard Chevalley-Serre basis of $\mathfrak{g}$ is given
by $\{e_{\beta},\, f_{-\beta}\,|\,\beta\in\Phi^{+}\}\cup\{h_{1},\ldots,h_{n}\}$.

Let the quantized universal enveloping algebra $\Uq$ be generated over $\mathbf{C}(q)$ by the elements $x_{i},\; y_{i},\; t_{i}^{\pm1}$, that correspond to the standard Chevalley-Serre basis. 
The algebra $\Uq$ becomes a Hopf algebra $H$ when equipped with 
the coproduct $\Delta$, counit $\epsilon$ and antipode $\sigma$ given by%
\footnote{Note that $t_{{\rm {Here}}}^{\pm1}=t_{{\scriptstyle {\rm {Letzter}}}}^{\mp1}$.
This is due to the consistency with (\ref{copEF}). %
}
\begin{align}
\Delta(x_{i}) & =x_{i}\otimes1+t_{i}^{-1}\otimes x_{i}, &  & \epsilon(x_{i})=0, &  & \sigma(x_{i})=-t_{i}x_{i},\el
\Delta(y_{i}) & =y_{i}\otimes t_{i}+1\otimes y_{i}, &  & \epsilon(y_{i})=0, &  & \sigma(y_{i})=-y_{i}t_{i}^{-1},\el
\Delta(t_{i}) & =t_{i}\otimes t_{i}, &  & \epsilon(t_{i})=1, &  & \sigma(t_{i})=t_{i}^{-1}.
\end{align}
Being a Hopf algebra, $\Uq$ admits left and right adjoint actions making
$\Uq$ into a left and right module. The (twisted) adjoint action is defined
as
\begin{align}
\left(\mbox{ad}\,x_{i}\right)b & =x_{i}b-(-1)^{[i][b]}t_{i}^{-1}bt_{i}x_{i}, &  & (\mbox{ad}_{r}\,x_{i})b=t_{i}bx_{i}-(-1)^{[i][b]}t_{i}x_{i}b,\el
\left(\mbox{ad}\,y_{i}\right)b & =y_{i}bt_{i}^{-1}-(-1)^{[i][b]}by_{i}t_{i}^{-1}, &  & (\mbox{ad}_{r}\,y_{i})b=by_{i}-(-1)^{[i][b]}y_{i}t_{i}^{-1}bt_{i},\el
\left(\mbox{ad}\,t_{i}\right)b & =t_{i}^{-1}bt_{i}, &  & (\mbox{ad}_{r}\,t_{i})b=t_{i}bt_{i}^{-1},
\end{align}
for all $b\in \Uq$. Here $(-1)^{[i][b]}$ represent the grading factor
of supercharges. We shall also be using the shorthand notation $\mbox{ad}\, y_{i_{1}}\!\cdots y_{i_{l}}=\mbox{ad}\, y_{i_{1}}\!\cdots\mbox{ad}\, y_{i_{l}}$
and similarly for $\mbox{ad}_{r}$. Finally, let us introduce the abelian
subgroup $T$ of $\Uq$ generated by the elements $t_{i}^{\pm1}$. Let $Q(\pi)$ be equal to the integral lattice generated by $\pi$, i.e  $Q(\pi) = \sum_{1\leq i\leq n}\mathbb{Z}\alpha_{i}$ and define $Q^{+}(\pi)=\sum_{1\leq i\leq n}\mathbb{N}\alpha_{i}$; here $\mathbb{N}$ is the set of non-negative integers. Then there is an isomorphism $\tau$ of abelian groups from $Q(\pi)$ to $T$ defined by $\tau(\alpha_{i})=t_{i}$, thus for every $\lambda\in\Phi$ there is an image $\tau(\lambda)\in T$.


\paragraph{Coideal subalgebra.}

A vector subspace $I$ of the Hopf algebra $H$ is called a left coideal if
\begin{equation}
\Delta(I)\subset H\otimes I.
\end{equation}
In the same way the right coideal may be defined, $\Delta(I)\subset I\otimes H$.
Let $M$ be a Hopf subalgebra of $H$ such that $\Delta\mathcal{M}\subset\mathcal{M}\otimes\mathcal{M}$.
Then $\left(\mbox{ad}\mathcal{M}\right)I$ (resp. $\left(\mbox{ad}_{r}\mathcal{M}\right)I$)
is an $\left(\mbox{ad}\mathcal{M}\right)$ (resp. $\left(\mbox{ad}_{r}\mathcal{M}\right)$)
invariant coideal of $H$. 

We shall be considering the scattering off the right boundary; thus we shall be interested in left coideals
only. However all considerations we shall present may be straightforwardly
extended for right coideals (scattering off the left boundary). 


\paragraph{Quantum symmetric pairs.}

Let $\theta$ be a non-trivial involution of $\mathfrak{g}$. It defines
a symmetric pair $(\mathfrak{g},\mathfrak{g}^{\theta})$, where $\mathfrak{g}^{\theta}$
is the $\theta$--fixed subalgebra of $\mathfrak{g}$. We shall assume
that $\theta$ is maximally split with respect to the Cartan algebra,
i.e.\ it satisfies the following three conditions%
\footnote{If $\theta$ is not maximally split it is often possible to replace
$\theta$ by a conjugate $\theta'=\psi\theta\psi^{-1}$ which is maximally
split; here $\psi$ is an automorphism of $\mathfrak{g}$. As an example, let 
 $\mathfrak{g}=\mathfrak{sl}(2)$. Then there is
only one non-trivial involution $\theta(h)=h$, $\theta(e)=-e$, $\theta(f)=-f$
leading to $\mathfrak{g}^{\theta}=\{h\}$. However it is isomorphic
to a maximally split involution $\theta'(h)=-h$, $\theta'(e)=-f$,
$\theta'(f)=-e$ with $\mathfrak{g}^{\theta'}=\{e-f\}$.}
\begin{eqnarray}
a) &  & \theta(\mathfrak{h})=\mathfrak{h},\el
b) &  & \mbox{if }\theta(h_{i})=h_{i},\mbox{ then }\theta(e_{i})=e_{i}\mbox{ and }\theta(f_{i})=f_{i},\el
c) &  & \mbox{if }\theta(h_{i})\neq h_{i},\mbox{ then }\theta(e_{i})\in\mathfrak{n}^{-}\mbox{ and }\theta(f_{i})\in\mathfrak{n}^{+}.\label{sp}
\end{eqnarray}

Quantum symmetric pairs are the quantum analogs of the pair of enveloping algebras $\Uq$, $U(\mathfrak{g}^{\theta})$ and consists of a pair of algebras $\Uq$ and $\mathcal{B}$, where that $\mathcal{B}$ is a left coideal subalgebra 
\begin{equation}
\Delta\mathcal{B}\subset \Uq\otimes\mathcal{B}.\label{B_coideal}
\end{equation}
The pair $(\mathcal{B},\Uq)$ is then called a quantum symmetric pair if $\mathcal{B}$ is the unique{\footnote{In the sense of \cite{L3}.}} maximal coideal subalgebra which specialized to $U(\mathfrak{g}^\theta)$ as $q\rightarrow 1$.

\paragraph{Construction of $\mathcal{B}$.} 
The explicit construction of $\mathcal{B}$ is given in \cite{L3}. It is formulated in terms of the root vectors of $\g$.

The involution $\theta$ of $\mathfrak{g}$ has an associated automorphism
$\Theta$ on the root system $\Phi$. Let $\pi_{\Theta}=\{\Theta(\alpha_i)=\alpha_i\,|\,\alpha_i\in\pi\}=\Theta(\pi)\cap\pi$, 
then (\ref{sp}) tells that $\Theta(-\alpha_{i})\in\Phi^{+}$ for all $\alpha_{i}\notin\pi_{\Theta}$.
More precisely,
\begin{align}\label{rootaut}
\Theta(-\alpha_{i})\in\sum_{\alpha_{j}\notin\pi_{\Theta}}\mathbb{N}\alpha_{j}+Q^{+}(\pi_{\Theta}), 
\qquad\text{for}\qquad \forall\alpha_{i}\notin\pi_{\Theta}. 
\end{align}
This implies that there
exists a permutation $p$ on the set $\{i\,|\,\alpha_{i}\in\pi\backslash\pi_{\Theta}\}$
such that $\forall\alpha_{i}\in\pi\backslash\pi_{\Theta} \Rightarrow \Theta(-\alpha_{i})-\alpha_{p(i)}\in Q^{+}(\pi_{\Theta})$.
Let $\pi^{*}$ be a maximal subset of $\pi\backslash\pi_{\Theta}$
such that $\alpha_{i}\in\pi^{*}$ if $i=p(i)$ or $\alpha_{p(i)}\notin\pi^{*}$.
Then for given $i$ such that $\alpha_{i}\in\pi^{*}$ there exists
a sequence $\{\alpha_{i_{1}},\ldots,\alpha_{i_{r}}\,|\,\forall\alpha_{i}\in\pi_{\Theta}\}$
and a set of positive integers $m_{1},\ldots,m_{r}$ such that the
involution $\theta$ may be lifted to the quantum involution $\tilde{\theta}$
of $\Uq$ subject to the following properties
\begin{eqnarray}
a) &  & \tilde{\theta}(q)=q^{-1},\el
b) &  & \tilde{\theta}(\tau(\lambda))=\tau(-\Theta(\lambda))\;\mbox{ for all }\;\tau(\lambda)\in T,\el
c) &  & \tilde{\theta}(x_{i})=x_{i}\;\mbox{ and }\;\tilde{\theta}(y_{i})=y_{i}\;\mbox{ for all }\;\alpha_{i}\in\pi_{\Theta},\el
d) &  & \tilde{\theta}(y_{i})=\bigl(\mbox{ad}_{r}x_{i_{1}}^{(m_{1})}\!\cdots x_{i_{r}}^{(m_{r})}\bigr)x'_{p(i)}\el
 &  & \mbox{and }\;\tilde{\theta}(y_{p(i)})=(-1)^{m(i)}\bigl(\mbox{ad}_{r}x_{i_{r}}^{(m_{r})}\!\cdots x_{i_{1}}^{(m_{1})}\bigr)x_{i}'\;\mbox{ for all }\;\alpha_{i}\in\pi^{*},\el
e) &  & \tilde{\theta}(x_{i}')=\bigl(\mbox{ad}_{r}y_{i_{1}}^{(m_{1})}\!\cdots y_{i_{r}}^{(m_{r})}\bigr)y_{p(i)}\el
 &  & \mbox{and }\;\tilde{\theta}(x_{p(i)}')=(-1)^{m(i)}\bigl(\mbox{ad}_{r}y_{i_{r}}^{(m_{r})}\!\cdots y_{i_{1}}^{(m_{1})}\bigr)y_{p(i)}\;\mbox{ for all }\;\alpha_{i}\in\pi^{*}.
\end{eqnarray}
Here $x_{i}'=t_{i}x_{i}$ and $x_{i}^{(m)}=x_{i}^{m}/[m]_{q_{i}}!\,$,
$y_{i}^{(m)}=y_{i}^{m}/[m]_{q_{i}}!\,$, where $q_{i}=q^{(\alpha_{i},\alpha_{i})/2}$.

Next, define a set $\mathcal{D}=\{\alpha_{i}\in\pi^{*}|\; i\neq p(i)\;\mbox{and}\;(\alpha_{i},\Theta(\alpha_{i}))\neq0\}$ and
let $\mathcal{M}$ denote the subalgebra of $\Uq$ generated by $\{x_{i},\; y_{i},\; t_{i}^{\pm1}\;|\;\alpha_{i}\in\pi_{\Theta}\}$. 
This setting allows us to define the coideal subalgebra $\mathcal{B}$ of $\Uq$
generated by $\mathcal{M}$, $T_{\Theta}=\{\tau(\lambda)|\,\Theta(\lambda)=\lambda\}$
and the set of twisted charges $B=\{B_{y,i},\; B_{y,i}\;|\;\alpha_{i}\in\pi\backslash\pi_{\Theta}\}$,
defined by
\begin{equation}
B_{y,i}=y_{i}t_{i}^{-1}+d_{y,i}\,\tilde{\theta}(y_{i})t_{i}^{-1}\qquad\mbox{and}\qquad B_{x,i}=t_{i}^{-1}x_{i}'+d_{x,i}\, t_{i}^{-1}\tilde{\theta}(x_{i}'),\label{twistB}
\end{equation}
with $d_{y,i},\, d_{x,i}=1$ for $\alpha_{i}\notin\mathcal{D}$ and
$d_{y,i},\, d_{x,i}\in\mathbf{C}$ otherwise. Note that $B_{x,i}$
may be obtained from $B_{y,i}$ with the help of the anti-automorphism
$\kappa$ of $\Uq$ which is defined as $\kappa(x_{i})=y_{i}t_{i}^{-1}$,
$\kappa(y_{i})=t_{i}x_{i}$, $\kappa(t_{i}^{\pm1})=t_{i}^{\pm1}$
and also if $a\in\mathbf{C}$ and $\bar{a}$ is a complex conjugate
of $a$, then $\kappa(au)=\bar{a}u$ for $\forall u\in \Uq$, i.e. $\kappa$
is a conjugate linear map and gives $\Uq$ a structure of Hopf $*$--algebra.
Then it is easy to see that $\kappa\left(B_{x,i}\right) = B_{y,i}$,
\begin{equation}
\kappa(t_{i}^{-1}x_{i}')=y_{i}t_{i}^{-1}\qquad\mbox{and}\qquad
 \kappa\bigl(d_{x,i}\,t_{i}^{-1}\tilde{\theta}(x_{i}')\bigr)=\bar{d}_{x,i}\,\tilde{\theta}(y_{i})t_{i}^{-1}.
\end{equation}
Furthermore, it implies that 
\begin{equation}
d_{y,i}=\bar{d}_{x,i}\,.\label{d_constrain}
\end{equation}


\subsection{Construction of the coideal subalgebra}

Having all the algebraic structures presented we are ready to explicitly construct the quantum affine coideal subalgebra $\widehat{\mathcal{B}}$ of $\widehat{\mathcal{Q}}$ by generalizing the results derived in the previous section. Let us start from inspecting the charges of $\widehat{\mathcal{Q}}$. It has eight regular supercharges, namely 
\begin{equation}
F_{2},\; F_{21},\; F_{32},\; F_{321}\qquad\mbox{and}\qquad E_{2},\; E_{21},\; E_{32},\; E_{321},\label{sup}
\end{equation}
where we have used a shorthand notation $F_{ijk}=[F_{i},[F_{j},F_{k}]]$ and the same for $E_{ijk}$. By replacing $F_{2}\to F_{4}$ and $E_{2}\to E_{4}$ eight affine supercharges are obtained,
\begin{equation}
F_{4},\; F_{41},\; F_{34},\; F_{341}\qquad\mbox{and}\qquad E_{4},\; E_{41},\; E_{34},\; E_{341}.\label{supaff}
\end{equation}
The replacement of the same type applied to (\ref{C123}) produces affine
partners of the central charges, $\hat{C}_{1}$, $\hat{C}_{2}$ and
$\hat{C}_{3}$. And finally, the affine partners of $F_{1}$, $E_{1}$
and $F_{3}$, $E_{3}$ are
\begin{equation}
\hat{F}_{1}=E_{432},\;\hat{E}_{1}=F_{432}\qquad\mbox{and}\qquad\hat{F}_{3}=E_{421},\;\hat{E}_{3}=F_{421}.
\end{equation}
 
The boundary we are considering does not respect bosonic symmetries
$E_{1}$, $F_{1}$, central charges $C_{2}$, $C_{3}$ and affine
charges $E_{4}$, $F_{4}$ (let us name these charges as the \textit{broken},
while the rest will be named as \textit{preserved}), thus it breakes
exactly half of the supercharges (\ref{sup}) and (\ref{supaff})
with the broken regular supercharges being 
\begin{equation}
E_{21},\; E_{321}\qquad\mbox{and}\qquad F_{21},\; F_{321}.
\end{equation}
In other words, we consider the involution that simply acts like 
\begin{align}
&\theta(X)= X, && \forall X\in \{E_2,E_3,F_2,F_3,K_1,K_2,K_3,K_4,C_1\} \el
&\theta(X)= -X, && \forall X\in \{E_1,F_1, E_4,F_4, C_2,C_3\}.
\end{align}
In the $q\to1$ limit this clearly gives rise to the symmetric pair 
\begin{eqnarray}
\mathfrak{g}^{\theta} & = & \{\mathfrak{H}_{1},\,\mathfrak{E}_{2},\,\mathfrak{F}_{2},\,\mathfrak{H}_{2},\,\mathfrak{E}_{3},\,\mathfrak{F}_{3},\,\mathfrak{H}_{3},\,\mathcal{U}_{2},\,\mathfrak{C}_{1}\},\el
\mathfrak{g}\backslash\mathfrak{g}^{\theta} & = & \{\mathfrak{E}_{1},\,\mathfrak{F}_{1},\,\mathfrak{C}_{2},\,\mathfrak{C}_{3}\},\label{qdecomp}
\end{eqnarray}
of the centrally extended $\mathfrak{sl}(2|2)$ algebra (see Appendix \ref{appMap}
for details on $q\to1$ limit).

Furthermore, it is easy to see that $\theta$ is isomorphic to a maximally split symmetric pair.
Thus the construction presented in section \ref{sub:Coideals} and the relation 
to the algebraic structures of the $Y=0$ giant graviton implies that for each broken regular charge, the 
algebra $\widehat{\mathcal{B}}$ must possess a corresponding twisted
affine charge satisfying coideal property \eqref{B_coideal}. 
We shall denote these charges as 
\begin{equation}
B=\{\widetilde{F}_{1},\;\widetilde{F}_{21},\;\widetilde{F}_{321},\;\widetilde{E}_{1},\;\widetilde{E}_{21},\;\widetilde{E}_{321},\;\widetilde{C}_{2},\;\widetilde{C}_{3}\}.
\end{equation}


\paragraph{Coideal subalgebra.}

The set of positive simple roots of $\widehat{\mathcal{Q}}$ is $\pi=\{\alpha_{1},\,\alpha_{2},\,\alpha_{3},\,\alpha_{4}\}$.
The boundary conditions imply that the corresponding root space automorphism $\Theta$ \eqref{rootaut}
acts on the simple roots as 
\begin{align}\label{auto}
\Theta(\alpha_{2}) &=\alpha_{2}, & \Theta(\alpha_{1})&=-\alpha_2-\alpha_3-\alpha_4,\el
\Theta(\alpha_{3}) &=\alpha_{3}, & \Theta(\alpha_{4})&=-\alpha_1-\alpha_2-\alpha_3.
\end{align}
Thus $\pi_{\Theta}=\{\alpha_{2},\,\alpha_{3}\}$
and it gives rise to a subalgebra $\mathcal{M}$ of $\widehat{\mathcal{Q}}$.
Note that $\alpha_4 = \delta - \bar{\theta}$ is the affine root where $\bar{\theta} = \alpha_1 + \alpha_2 + \alpha_3$ is the highest root of the non-affine algebra $\mathcal{Q}$. However we are interested in the finite dimensional representations which are constructed by dropping all imaginary roots; thus giving the constraint $K_1 K_2 K_3 K_4 = 1$ \cite{BGM,LMR}.
 
We shall build $\widehat{\mathcal{B}}$ based on the affine extension,
hence we set $\pi^{*}=\{\alpha_{4}\}$. This fixes the permutation
map $p$ to act as $p(4)=1$. Next, with the help of \eqref{twistB} and \eqref{auto}, we define the
twisted affine charges to be
\footnote{Equivalently, one could choose $\pi^{*}=\{\alpha_1\}$ as a starting point giving $\widetilde{E}_1 = E'_1 K_{1}^{-1} + d_{x}\,\tilde{\theta}(E'_{1})K_{1}^{-1} $ and $\widetilde{F}_1 = F_1 K_{1}^{-1} + d_{y}\,\tilde{\theta}(F_{1})K_{1}^{-1} $ where $\tilde{\theta}(E'_{1})=\left(\mbox{ad}_{r}F_{2}\,\mbox{ad}_{r}F_{3}\right)F_{4}$ and $\tilde{\theta}(F_{1})=\left(\mbox{ad}_{r}E_{2}\,\mbox{ad}_{r}E_{3}\right)E'_{4}$. }
\begin{align}
 & \widetilde{E}_{321}=F_{4}K_{4}^{-1}+d_{y}\,\tilde{\theta}(F_{4})K_{4}^{-1}, &  & \tilde{\theta}(F_{4})=\left(\mbox{ad}_{r}E_{3}\,\mbox{ad}_{r}E_{2}\right)E_{1}',\label{twE321}\\
 & \widetilde{F}_{321}=E_{4}'K_{4}^{-1}+d_{x}\,\tilde{\theta}(E_{4}')K_{4}^{-1}, &  & \tilde{\theta}(E_{4}')=\left(\mbox{ad}_{r}F_{3}\,\mbox{ad}_{r}F_{2}\right)F_{1},\label{twF321}
\end{align}
Then with the help of the right adjoint action $\mbox{ad}_{r}\mathcal{M}$
we construct the rest of the twisted affine charges,
\begin{align}
\widetilde{E}_{21} & =\left(\mbox{ad}_{r}F_{3}\right)\widetilde{E}_{321}, & \widetilde{F}_{21} & =\left(\mbox{ad}_{r}E_{3}\right)\widetilde{F}_{321}, \label{twEF21}\\
\widetilde{E}_{1} & =\left(\mbox{ad}_{r}F_{2}\,\mbox{ad}_{r}F_{3}\right)\widetilde{E}_{321}, & \widetilde{F}_{1} & =\left(\mbox{ad}_{r}E_{2}\,\mbox{ad}_{r}E_{3}\right)\widetilde{F}_{321}, \label{twEF1}\\
\widetilde{C}_{2} & =\left(\mbox{ad}_{r}E_{2}\right)\widetilde{E}_{321}, & \widetilde{C}_{3} & =\left(\mbox{ad}_{r}F_{2}\right)\widetilde{F}_{321}.\label{twC23}
\end{align}
Let us show the coideal property for the these charges explicitly.
However it is enough to show this property for the charges (\ref{twE321})
and (\ref{twF321}) only, 
\begin{eqnarray}
\Delta\widetilde{E}_{321} & = & F_{4}K_{4}^{-1}\otimes1+UK_{4}^{-1}\otimes\widetilde{E}_{321}+d_{y}\,\tilde{\theta}(F_{4})K_{4}^{-1}\otimes K_{5}\el
 &  & +d_{y}(q^2-1)\left(q^{-1}K_{4}^{-1}\left(\mbox{ad}_{r}E_{2}\right)E_{1}'\otimes K_{5}E_{3}
- U E_{1}'K_{4}^{-1}\otimes K_1K_4^{-1}\left(\mbox{ad}_{r}E_{3}\right)E_{2}'\right)\el
 & \in & \widehat{\mathcal{Q}}\otimes\widehat{\mathcal{B}},
\end{eqnarray}
and
\begin{eqnarray}
\Delta\widetilde{F}_{321} 
&=& E_{4}'K_{4}^{-1}\otimes1+U^{-1}K_{4}^{-1}\otimes\widetilde{F}_{321}+d_{x}\,\tilde{\theta}(E'_4)K_{4}^{-1}\otimes K_{5}\el
&& -d_{x}(q^{2}-1)\left(K_{4}^{-1}\left(\mbox{ad}_{r}F_{2}\right)F_{1}\otimes K_{3}^{-1}K_{5}F_{3}
-U^{-1}K_{4}^{-1}F_{1}\otimes\left(\mbox{ad}_{r}F_{3}\right)F_{2}K_{1}K_{4}^{-1}\right)\el
 & \in & \widehat{\mathcal{Q}}\otimes\widehat{\mathcal{B}}.
\end{eqnarray}
Here $K_{5}=K_{1}K_{2}K_{3}K_{4}^{-1}$ and the coideal property is
satisfied provided $K_{1}K_{4}^{-1}\in T_{\Theta}$.  
This is easy to see, because the linear combination $\alpha_{1}-\alpha_{4}$ is invariant under the automorphism \eqref{auto}
as $\Theta(\alpha_{1}-\alpha_{4})=\alpha_{1}-\alpha_{4}$.
The coideal property for the rest of the charges, \eqref{twEF21}, \eqref{twEF1} and \eqref{twC23}, is obvious since
$\widehat{\mathcal{B}}$ is invariant under the adjoint action of
$\mathcal{M}$. 


\paragraph{Reflection algebra.}

In order to build the boundary scattering theory we need to have representations
of $\widehat{\mathcal{Q}}$ corresponding to incoming and reflected states.
We shall be bearing on the reflection Hopf algebra constructed in \cite{MR2}. 

Let the representation defined in section \ref{sec:Q} describe incoming states carrying momentum $p$. It is related to the deformation parameter as $U=e^{ip}$. Then the representation corresponding to the reflected states with momentum $-p$ shall have deformation parameter equal to $e^{-ip}=U^{-1}$. Next, the total fermion and boson number conservation together with the energy conservation constrains central elements $V$ and $K_i$ to be invariant under the reflection.

This implies that there is a reflection automorphism $\kappa$ of the algebra defined as
\begin{equation}
\kappa:(V,U)\mapsto(\underline{V},\underline{U}) \qquad\mbox{and}\qquad \kappa:(E_{j},F_{j},K_{j})\mapsto(\underline{E}_{j},\underline{F}_{j},\underline{K}_{j}),
\end{equation}
where the underlined charges describe the representation of reflected states and the constraints
\begin{equation}
\underline{U}=U^{-1},\qquad \underline{V}=V, \qquad \underline{K}_i=K_i
\end{equation}
define the representation uniquely. The representation labels $\underline{a},\,\underline{b}\,,\underline{c},\,\underline{d}$
associated to the charges $\underline{E}_{j},\underline{F}_{j}$ may
be obtained from (\ref{rep}) by replacing $U\mapsto U^{-1}$ and
similarly for the affine ones. Then, we can express the labels of
the reflected charges in terms of the initial ones as
\begin{equation}
\underline{a}=\frac{\,\underline{\gamma}\,}{\gamma}a,\qquad\underline{b}=\frac{\gamma\alpha^{2}}{\underline{\gamma}}\frac{cd}{a}V^{2},\qquad\underline{c}=\frac{\underline{\gamma}}{\gamma\alpha^{2}}\frac{ab}{d}V^{-2},\qquad\underline{d}=\frac{\,\gamma\,}{\underline{\gamma}}d,\label{ref_abcd}
\end{equation}
giving 
\begin{align}
\underline{a} &=\sqrt{\frac{g}{[M]_{q}}}\underline{\gamma}, && \underline{b}=\sqrt{\frac{g}{[M]_{q}}}\frac{\alpha}{\underline{\gamma}}\frac{\tilde{g}^{2}(x^{+}-x^{-})}{g^{2}(1+\xi x^{-})(\xi+x^{+})}, \el
\underline{c} & =\sqrt{\frac{g}{[M]_{q}}}\frac{\underline{\gamma}}{\alpha\, V}\frac{gq^{\frac{M}{2}}(\xi x^{-}+1)}{i\tilde{g}\, x^{-}}, && \underline{d}=\sqrt{\frac{g}{[M]_{q}}}\frac{\tilde{g}\, q^{\frac{M}{2}}V}{i\, g\,\underline{\gamma}}\frac{x^{+}-x^{-}}{\xi x^{+}+1}, \label{abcd_ref}
\end{align}
The extension to the affine case is obvious. Here we have chosen
$\underline{a}=\frac{\,\underline{\gamma}\,}{\gamma}a$ as an initial
constraint with $\underline{\gamma}$ being the reflected version
of $\gamma$, i.e.\ $\kappa(\gamma)=\underline{\gamma}$. By comparing 
\eqref{abcd_ref} with \eqref{abcd} we find the reflection map for the 
$x^\pm$ parameterization to be
\begin{equation}
\kappa:x^{\pm}\mapsto-\frac{x^{\mp}+\xi}{\xi x^{\mp}+1}.
\end{equation}
It is in agreement with the one conjectured in \cite{MN}%
\footnote{The authors of \cite{MN} are using the $x^{\pm}$ parametrization
of \cite{BK}, while we use the one of \cite{BGM}. The map between
these two is $x_{\text{\tiny{BK}}}^{\pm}=g{\tilde{g}}^{-1}(x_{\text{\tiny{BGM}}}^{\pm}+\xi)$.%
}.
In the $q\to1$ limit this maps reduces to the usual reflection
map $\kappa:x^{\pm}\mapsto-x^{\mp}$.

Let us also introduce the reflected coproducts of $E_i$ and $F_i$ associated to the reflection Hopf algebra \cite{MR2}. They are
\begin{equation}
\Delta^{\! ref\!}(E_{j})=\underline{E}_{j}\otimes1+K_{j}^{-1}U^{-\delta_{j,2}-\delta_{j,4}}\otimes E_{j},
 \quad \Delta^{\! ref\!}(F_{j})=\underline{F}_{j}\otimes K_{j}+U^{+\delta_{j,2}+\delta_{j,4}}\otimes F_{j}.
\end{equation}
These shall play an important role in finding the explicit form of the reflection matrix.

The expressions in \eqref{ref_abcd} may be casted in a matrix form
\begin{align}
\begin{pmatrix}\underline{a}&\underline{b}\\ \underline{c}&\underline{d} \end{pmatrix} D
=T\begin{pmatrix}a&b\\ c&d \end{pmatrix}T^{-1} 
\quad\text{with}\quad
D=\begin{pmatrix} \gamma/\underline{\gamma} & 0 \\ 0 & \underline{\gamma}/\gamma \end{pmatrix}, 
\quad
T=\begin{pmatrix} U^{-2} & 0 \\ 0 & -z \end{pmatrix}, 
\end{align}
revealing the explicit relation between two isomorphic representations of $\afQ$. Here $\gamma$ and $\underline{\gamma}$ are unconstrained parameters defining the representations of incoming and reflected states. Thus the matrix $D$ may be understood as a matrix relating two different basis, while the matrix $T$ expresses the reflection automorphism of the algebra. Indeed, physical intuition tells that the representations of incoming and reflected states should be related via the spectral and deformation parameters only.

Finally, we want to perform some checks of our constructions. Firstly, the twisted affine central
charges $\widetilde{C}_{2}$ and $\widetilde{C}_{3}$ (\ref{twC23}) must be conserved under the reflection.
Thus requiring $\underline{\widetilde{C}}_{2}=\widetilde{C}_{2}$ and $\underline{\widetilde{C}}_{3}=\widetilde{C}_{3}$
we find 
\begin{equation}
d_{y}=\frac{\tilde{g}}{g\alpha\tilde{\alpha}}\qquad\mbox{and}\qquad d_{x}=-\alpha\tilde{\alpha}\frac{\tilde{g}}{g}\,.
\end{equation}
Let us make a direct link to the constraint (\ref{d_constrain})
arising from the Hopf $*$--algebra.  
Requiring $\tilde{g}/g$ to be real, we find $(\alpha\tilde{\alpha})^{2}=-1$ having a solution $\tilde{\alpha}=1$
and $\alpha=i$ which corresponds to the usual setting of unitary
representations.

Secondly, the spectral parameter $z$ associated the algebra is required transform as $\kappa: z \mapsto z^{-1}$ under the reflection map. This is indeed true and follows straightforwardly when applying map $\kappa$ to \eqref{z}.  


\paragraph{Yangian limit.}

The algebra $\afQ$ in the $q\to1$ limit has no singular elements and the naive $q\to1$ limit leads to the undeformed universal enveloping algebra. The relation to the associated Yangian algebra was explicitly shown in \cite{BGM} by considering the the specific combinations of charges of $\afQ$ that are singular in the $q\to1$ limit. The construction presented in \cite{BGM} is very closely related to the so-called Drinfeldian \cite{Tolstoy}. However the twisted affine charges (\ref{twE321}\,-\,\ref{twC23}) are already of the required form. Thus the algebra $\afB$ in the rational $q\to1$ limit is isomorphic to the associated twisted Yangian \eqref{twistY} proposed by \cite{AN,MR1}.
The explicit relations between the quantum affine and Yangian charges are
\begin{align}
& \underset{q\to1}{\lim}\;\frac{\alpha\tilde{\alpha}\widetilde{E}_{321}}{2(q-1)} = -\widetilde{\mathfrak{E}}_{321} - g\alpha\,\mathfrak{F}_{2}
&& \underset{q\to1}{\lim}\;\frac{\widetilde{F}_{321}}{2\alpha\tilde{\alpha}(q-1)} = -\widetilde{\mathfrak{F}}_{321} - \frac{g}{\alpha}\mathfrak{E}_{2},
\el
& \underset{q\to1}{\lim}\;\frac{\alpha\tilde{\alpha}\widetilde{E}_{21}}{2(q-1)} = \widetilde{\mathfrak{E}}_{21}+g\alpha\,\mathfrak{F}_{23}, 
&& \underset{q\to1}{\lim}\;\frac{\widetilde{F}_{21}}{2\alpha\tilde{\alpha}(q-1)} = \widetilde{\mathfrak{F}}_{21}-\frac{g}{\alpha}\mathfrak{E}_{23},
\el
& \underset{q\to1}{\lim}\;\frac{\alpha\tilde{\alpha}\widetilde{E}_{1}}{2(q-1)} = \widetilde{\mathfrak{E}}_{1}, 
&& \underset{q\to1}{\lim}\;\frac{\widetilde{F}_{1}}{2\alpha\tilde{\alpha}(q-1)} = -\widetilde{\mathfrak{F}}_{1},
\el
& \underset{q\to1}{\lim}\;\frac{\alpha\tilde{\alpha}\widetilde{C}_{2}}{2(q-1)} = -\widetilde{\mathfrak{P}}+g\alpha\,\mathfrak{H}_{2}, 
&& \underset{q\to1}{\lim}\;\frac{\widetilde{C}_{3}}{2\alpha\tilde{\alpha}(q-1)} = -\widetilde{\mathfrak{K}}-\frac{g}{\alpha}\mathfrak{H}_{2}. 
\label{RLim}
\end{align}
We have also checked that these relations hold at both algebra and coalgebra level. 
Noting that the contributions of the extended central charges for the twisted Yangian generators are 
also recovered in the above limit(see also appendix \ref{appYD3}).


\section{Boundary scattering}

In this section we consider the boundary scattering theory for the deformed Hubbard model and find the explicit form of the bound state reflection matrix. Moreover, we explicitly solve the reflection equation and show that the reflection matrix $\mathbb{K}$ is indeed invariant under the coideal subalgebra $\afB$.

\paragraph{Reflection matrix.}

The boundary we are considering is a singlet with respect to the boundary algebra $\afB$, thus it may be represented via the boundary vacuum state $|0\rangle_{B}$. It is annihilated by all charges of it, with the exception of the generators $K_{i}$, which actually keep the boundary invariant,
\begin{equation}
K_{i}\,|0\rangle_{B} = q^{H_{i}}\,|0\rangle_{B} = |0\rangle_{B}.
\end{equation}
We define the reflection matrix to be the intertwining matrix 
\begin{equation}
\K\,|m,n,k,l\rangle\otimes|0\rangle_{B}=K_{(m,n,k,l)}^{(a,b,c,d)}\,|a,b,c,d\rangle\otimes|0\rangle_{B}.\label{K}
\end{equation}
The space of states $|m,n,k,l\rangle$ is $4M$-dimensional and can be decomposed into four $4M=(M+1)+(M-1)+M+M$ subspaces that have the orthogonal basis 
\begin{align}
|k\rangle^{1} & =|0,0,k,M\!-\! k\rangle, &  & k=0\ldots M,\nonumber \\
|k\rangle^{2} & =|1,1,k\!-\!1,M\!-\! k\!-\!1\rangle, &  & k=1\ldots M-1,\nonumber \\
|k\rangle^{3} & =|1,0,k,M\!-\! k\!-\!1\rangle, &  & k=0\ldots M-1,\nonumber \\
|k\rangle^{4} & =|0,1,k,M\!-\! k\!-\!1\rangle, &  & k=0\ldots M-1.\label{basis}
\end{align}


\paragraph{Symmetry constraints.}

The reflection matrix (\ref{K}) is required to be invariant under the coproducts of the boundary algebra \cite{MR1} 
\begin{equation}
\K\,\Delta(J)-\Delta^{\! ref\!}(J)\,\K=0, \qquad \forall J\in\widehat{\mathcal{B}}. \label{Kinv}
\end{equation}
The form of reflection matrix is constrained by the bosonic charges $E_{3}$ and $F_{3}$ to five independent sets of coefficients
\begin{eqnarray}
\K\,|k\rangle^{1} & = & A_{k}\,|k\rangle^{1}+D_{k}\,|k\rangle^{2},\nonumber \\
\K\,|k\rangle^{2} & = & B_{k}\,|k\rangle^{2}+E_{k}\,|k\rangle^{1},\nonumber \\
\K\,|k\rangle^{\alpha} & = & C_{k}\,|k\rangle^{\alpha},
\end{eqnarray}
where $\alpha=3,\,4$ and we have dropped the boundary vacuum state. We note that the basis (\ref{basis}) was chosen is such a way that the reflection matrix would act diagonally on the quantum number $k$. Also we are working in an orthogonal, but not orthonormal basis in order to avoid having normalization factors appearing in explicit expressions. However switching to the orthonormal basis is rather easy and requires only extra factors of $\left([k]![M-k]!\right)^{\frac{1}{2}}$ and $\left([k]![M-k]!\right)^{-\frac{1}{2}}$ to be added to $D_{k}$ and $E_{k}$ respectively.

We start by determining the limiting conditions - the constraints for reflection coefficients $A_{0}$, $D_{0}$, $C_{0}$ and $A_{M}$, $D_{M}$, $C_{M}$. This can be achieved by considering reflection of the lowest state $|0\rangle^{1}$: 
\begin{equation}
\K\,|0\rangle^{1}=A_{0}\,|0\rangle^{1},\qquad\mbox{thus}\qquad D_{0}=0.\label{D0}
\end{equation}
Then the invariance condition (\ref{Kinv}) for the charge $E_{2}$,
\begin{equation}
\left(\K\, E_{2}-\underline{E}_{2}\,\K\right)|0\rangle^{1}=0,
\qquad\mbox{gives}\qquad C_{0}=\frac{\,\underline{a}\,}{a}A_{0}=\frac{\,\underline{\gamma}\,}{\gamma}A_{0}.
\end{equation}
We choose the overall normalization to be $A_{0}=1$. 
The same constraint may be found by considering the reflection of states $|0\rangle^{\alpha}$ 
and the charge $F_{2}$. Similar considerations for the highest state $|M\rangle^{1}$ give 
\begin{equation}
D_{M}=0\qquad\mbox{and}\qquad A_{M}=\frac{\, c\,}{\underline{c}}C_{M-1}
=-\frac{\gamma}{z\,U^2\,\underline{\gamma}}C_{M-1}.
\end{equation}

Next we turn to the states $|k\rangle^{\alpha}$ as they scatter from the boundary diagonally. The twisted affine charge $\widetilde{F}_{1}$ acts on these states as a raising operator 
\begin{align}
&\widetilde{F}_{1}|k\rangle^{\alpha} =f_k(z)|k+1\rangle^{\alpha},\qquad 
\underline{\widetilde{F}}_{1}|k\rangle^{\alpha} =f_k(1/z)|k+1\rangle^{\alpha},
\el
\text{with}\qquad 
&f_k(z)\equiv d_x[M-k-1]_{q}q^{-M/2-k-1}\left(q^{M}-q^{2k+2}\mathit{z}\right)V^{-1}\,.
\end{align}
The invariance condition then straightforwardly gives
\begin{equation}
C_{k+1}\, f_{k}(z)-f_{k}(1/z)\, C_{k}=0,
\end{equation}
leading to the iterative relation 
\begin{equation}
C_{k}=\frac{f_{k-1}(1/z)}{f_{k-1}(z)}C_{k-1}
=\frac{q^{M}-q^{2k}/z}{q^{M}-q^{2k}z}C_{k-1}.\label{Ck}
\end{equation}
This relation is then simply solved by
\begin{align}
C_k  = C_0 \prod_{n=1}^{k} \frac{q^{M}-q^{2n}/z}{q^{M}-q^{2n}z}.\label{Ck2}
\end{align}
The coefficients $C_{k}$ are (anti)symmetric up to a factor of $z$ under the interchange $k\to M-k-1$ for $M$ being (even)odd,
\begin{align}
\label{cov}
z^k C_{k} & = - z^{M-k-1} C_{M-k-1} \quad \text{for}\quad M=even \quad \text{and} \quad k=0,\,...\,,\,M/2-1, \el
z^k C_{k} & = z^{M-k-1} C_{M-k-1} \qquad \!\text{for}\quad M=odd \quad\;\, \text{and} \quad k=0,\,...\,,\,(M-1)/2-1.
\end{align}
This symmetry comes from the requirement that the reflection is covariant under the renaming of bosonic indices $1\leftrightarrow2$ as the reflection is of a diagonal type for the states $|k\rangle^{\alpha}$. However this is not the case for the states $|k\rangle^{1,2}$, thus there is no such symmetry for the rest of the reflection coefficients. 
The factors of $z$ in \eqref{cov} arises due to the non-commutative nature of the model. 
In the $q\to1$ limit this (anti)covariance specializes to (anti)symmetry for $M$ being (even)odd, as observed in \cite{LP}.

The remaining reflection coefficients, as we shall show, will be expressed in terms of $C_{k}$ and $C_{k-1}$. 
Requiring the reflection matrix to be invariant under the charges $E_{2}$ and $F_{2}$ on the bosonic states $|k\rangle^{1,2}$, 
we obtain the following set of separable equations 
\begin{align}
 & D_{k}\,\underline{b}-[M\!-\! k]_{q}\left(C_{k}\, a-A_{k}\,\underline{a}\right)=0, &  & C_{k}\, b-[M\!-\! k]_{q}E_{k}\,\underline{a}-B_{k}\,\underline{b}=0,\nonumber \\
 & D_{k}\,\underline{d}+[k]_{q}\left(C_{k-1}\, c-A_{k}\,\underline{c}\right)=0, &  & C_{k-1}\, d+[k]_{q}E_{k}\,\underline{c}-B_{k}\,\underline{d}=0,
\end{align}
with the unique solution 
\begin{align}\label{ABCDstart}
A_{k} & =\left(C_{k-1}[k]_{q}\underline{\mathit{b}}\mathit{c}+C_{k}[M\!-\! k]_{q}\mathit{a}\underline{\mathit{d}}\right)/N, &  & 
D_k =[k]_{q}[M\!-\! k]_{q}\left(C_{k}\mathit{a}\underline{\mathit{c}}-C_{k-1}\underline{\mathit{a}}\mathit{c}\right)/N, \el
B_k& =\left(C_{k}[k]_{q}\mathit{b}\underline{c}+C_{k-1}[M\!-\! k]_{q}\underline{a}\mathit{d}\right)/N, &  & 
E_{k}=\left(C_{k}\mathit{b}\underline{\mathit{d}}-C_{k-1}\underline{\mathit{b}}\mathit{d}\right)/N, 
\end{align}
where the normalization factor $N$ is 
\begin{align}
\label{Nq}
N=[k]_{q}\,\underline{\mathit{b}}\,\underline{\mathit{c}}+[M\!-\! k]_{q}\,\underline{\mathit{a}}\,\underline{\mathit{d}}
=\frac{Vq^{M/2-k}-V^{-1}q^{-M/2+k}}{q-q^{-1}}\,. 
\end{align}
Writing the coefficients explicitly in terms of $x^{\pm}$ parametrization we then finally obtain
\begin{align}
A_{k} & =\frac{\gamma\,\tilde{g}\, q^{\frac{M}{2}}\left(x^{-}-x^{+}\right)\left(\tilde{g}^{2}q^{M}[k]_{q}C_{k-1}
-g^{2}[M\!-\! k]_{q}C_{k}\left(\xi+x^{+}\right)^{2}\right)V}{i\underline{\gamma}g^{2}[M]_{q}\left(\xi+x^{+}\right)^{2}\left(1+\xi x^{+}\right)N},
\el
B_{k} & =\frac{i\underline{\gamma}\, q^{-\frac{M}{2}}\left(x^{-}-x^{+}\right)\left(\tilde{g}^{2}[M\!-\! k]_{q}C_{k-1}\left(x^{-}\right)^{2}
-g^2q^{M}[k]_{q}C_{k}\left(1+\xi x^{-}\right)^{2}\right)}{\gamma\tilde{g}[M]_{q}\left(x^{-}\right)^{2}\left(1+\xi x^{-}\right)V\, N},
\el
D_{k} & =\frac{\gamma\underline{\gamma}\, q^{\frac{M}{2}}[k]_{q}[M\!-\! k]_{q}\left(\tilde{g}^{2}C_{k-1}x^{-}+g^{2}C_{k}
\left(1+\xi x^{-}\right)\left(\xi+x^{+}\right)\right)}{i\alpha\tilde{g}[M]_{q}\, x^{-}\left(\xi+x^{+}\right)V\, N},
\el
E_{k} & =\frac{i\alpha\,\tilde{g}\, q^{\frac{M}{2}}\left(x^{-}-x^{+}\right)^{2}\left(\tilde{g}^{2}C_{k-1}x^{-}+g^{2}C_{k}
\left(1+\xi x^{-}\right)\left(\xi+x^{+}\right)\right)V}
{\gamma\underline{\gamma}\, g^{2}[M]_{q}\, x^{-}\left(1+\xi x^{-}\right)\left(\xi+x^{+}\right)\left(1+\xi x^{+}\right)N}.
\label{ABCDend}
\end{align}


\paragraph{Unitarity.} The reflection matrix satisfies the unitarity constraint
\begin{align}
\K(p)\,\K(-p)=1.
\end{align}


\paragraph{Rational limit.}

In the $q\to1$ limit the reflection coefficients get reduced to 
\begin{align}\label{ABCD1start}
A_{k} & =\frac{\,\gamma\,}{\underline{\gamma}}\frac{x^{-}}{x^{+}N}\left((M\!-\! k)C_{k}(x^{+})^{2}-kC_{k-1}\right), &  & B_{k}=\frac{\,\underline{\gamma}\,}{\gamma}\frac{x^{+}}{x^{-}N}\left((M\!-\! k)C_{k-1}(x^{-})^{2}-kC_{k}\right),\nonumber \\
D_{k} & =\frac{\gamma\underline{\gamma}\,}{\alpha}\frac{k(M\!-\! k)\left(C_{k}x^{+}+C_{k-1}x^{-}\right)}{N(x^{+}-x^{-})}, &  & E_{k}=\frac{\alpha}{\gamma\underline{\gamma}}\frac{x^{-}-x^{+}}{N}\left(C_{k}x^{+}+C_{k-1}x^{-}\right),
\end{align}
and the coefficients $C_k$ and the normalization $N$ are given by 
\footnote{%
Here we have rescaled the normalization factor as 
$N_{\eqref{Nq}}\to N_{\eqref{ABCD1end}}/(x^+x^--1)$ in $q\to1$ limit.}
\begin{align}
C_{k}=\frac{\;\;2igu-M+2k}{-2igu-M+2k}C_{k-1},\qquad 
&N=k+(M\!-\! k)x^{-}x^{+}. \label{ABCD1end}
\end{align}
By choosing $\gamma=\sqrt{i(x^{-}-x^{+})}$ and $\underline{\gamma}=\gamma$
these are in agreement with the ones found in \cite{LP}.%
\footnote{Up to some factors due to different choice of the basis \eqref{basis} with respect to the one in \cite{LP}.}


\paragraph{Fundamental representation.}

In this case $M=1$ and the state $|k\rangle^{2}$ is absent, thus the reflection matrix is purely diagonal. The charges $E_{2}$ and $F_{2}$ constrain the reflection coefficients to be 
\begin{align}
A_{0} & =\frac{\, a\,}{\underline{a}}C_{0}=\frac{\,\gamma\,}{\underline{\gamma}}C_{0},\\
A_{1} & =\frac{\, c\,}{\underline{c}}C_{0}=-\frac{\gamma}{\underline{\gamma}} \frac{C_0}{z\, U^2}\;\underset{q\to1}{\longrightarrow}\;-\frac{\,\gamma\,}{\underline{\gamma}}\frac{x^{-}}{x^{+}}C_{0}.
\end{align}
Again, choosing the normalization to be $A_{0}=1$ this is in agreement with \cite{MN} and with \cite{HM} in the rational limit.


\paragraph{Reflection Equation.}

In order to show the integrability of the model, we have to show that the reflection matrix is a solution of the reflection equation (boundary Yang-Baxter equation). In fact, we shall explicitly derive the coefficient $C_{k}$ by solving the reflection equation. The unique solution we find agrees perfectly with the coefficients that are derived from the symmetry considerations. This explicitly proves that $\mathbb{K}$ respects the symmetry coideal algebra $\mathcal{B}$.

Consider two states with bound state numbers $M_{1},\; M_{2}$ and spectral parameters $z_{1},\; z_{2}$. Let us denote $\K_{i}=\K_{M_{i}}(z_{i})$ and $\mathbb{S}_{ij}=\S_{M_{i}M_{j}}(z_{i},z_{j})$ and also let the underscored index indicate that the corresponding representation is reflected. Then the reflection equation is then given by 
\begin{align}
\K_{2}\mathbb{S}_{2\underline{1}}\K_{1}\mathbb{S}_{12}=\S_{\underline{2}\underline{1}}\K_{1}\mathbb{S}_{1\underline{2}}\K_{2},
\end{align}
which explicitly written out in components reads as
\begin{equation}
K_{\gamma}^{\delta}(z_{2})\, S_{\beta,c}^{\gamma,d}(z_{2},z^{-1}_{1})\, K_{b}^{c}(z_{1})\, S_{a,\alpha}^{b,\beta}(z_{1},z_{2}) = 
S_{\gamma,c}^{\delta,d}(z_{2}^{-1},z^{-1}_{1})\, K_{b}^{c}(z_{1})\, S_{a,\beta}^{b,\gamma}(z_{1},z^{-1}_{2})\, K_{\alpha}^{\beta}(z_{2}),
\end{equation}
where we have used Roman and Greek letter to distinguish indices of the fist and second states respectively. 

Let us first consider states of the form $|k_{1}\rangle^{\alpha}\otimes|k_{2}\rangle^{\alpha}$, because the reflection matrix acts diagonally on these states. This corresponds to the \textit{subspace I} case in terms of the analysis performed in \cite{LMR}. Then the reflection equation becomes 
\begin{align}
 & \sum_{n=0}^{k_{1}+k_{2}}C_{m}(z_{1})\mathscr{X}_{m}^{K-n,n}(z_{2},z_1^{-1})C_{n}(z_{2})\mathscr{X}_{n}^{k_{1},k_{2}}(z_{1},z_{2})=\nonumber \\
 & \qquad\qquad\qquad\sum_{n=0}^{k_{1}+k_{2}}\mathscr{X}_{m}^{n,K-n}(z^{-1}_{2},z_{1}^{-1})C_{n}(z_{1})\mathscr{X}_{n}^{k_{1},k_{2}}(z_{1},z_{2}^{-1})C_{k_{2}}(z_{2}).
\end{align}
We will now proceed with the derivation of $C_{k}$. For $k_{1}=k_{2}=0$ we easily find that the reflection equation is satisfied. Next we consider the state where $k_{1}=1,k_{2}=0$. In this case the reflection equation is satisfied provided that $C_{1}$ satisfies the following relation 
\begin{align}
C_{1}(z_{1})=C_{0}(z_{1})\left[1-\frac{z_{1}^{2}-1}{z_{1}\left(\frac{q^{M_{1}-M_{2}}\left(z_{2}^{2}C_{1}(z_{2})-C_{0}(z_{2})\right)}{z_{2}(C_{1}(z_{2})-C_{0}(z_{2}))}+z_{1}\right)}\right].
\end{align}
The right hand side is allowed to depend solely on $z_{1}$ thus there are two solutions, a trivial one $C_{1}=C_{0}$ and 
\begin{align}
C_{1}=\frac{z^{-1}-A\,q^{M}}{z-A\,q^{M}}C_{0}.
\end{align}
The latter solution has an undetermined constant $A$. This coefficient may be determined by either considering the rational limit, or by studying the reflection equation involving states from the {\textit{subspace II}} of \cite{LMR}. Both arguments lead to $A = q^{-2}$. Finally, by studying a state with $k_{1}=2,\; k_{2}=0$ we can solve for $C_{2}$ and so on. This leads to the following solution
\begin{align}
C_{k}=C_{0}\prod_{i=1}^{k}\frac{q^{M}-q^{2i}/z}{q^{M}-q^{2i}z},
\end{align}
which perfectly agrees with (\ref{Ck2}). As expected, the trivial solution does not solve the reflection equation in general case.

Subsequently we have numerically checked the reflection equation for generic values of $k_1,k_2,n,M_1,M_2$ for all different states from \textit{subspace II} and \textit{subspace III} and found it to be satisfied.


\section{Discussion}

In this work we have considered open boundary conditions for the deformed Hubbard model of the same type as for the $Y=0$ giant graviton in the AdS/CFT correspondence. In this situation exactly half of the supersymmetries are broken. The symmetry algebra compatible with these boundary conditions is a twisted coideal quantum affine algebra $\afB$ of the quantum affine deformed algebra $\afQ$.

Inspired by results obtained for semisimple Lie algebras \cite{L1,L2,L3,L4} we provide an explicit construction of $\afB$ that is of a coideal form. We then find the corresponding reflection matrix for arbitrary bound states and show that it satisfies the reflection equation (boundary Yang-Baxter equation). Conversely, we explicitly solve the reflection equation and find that our reflection matrix corresponds to the unique solution compatible with the boundary conditions. This proves that the reflection matrix indeed respects $\afB$ as symmetry algebra. However the boundary algebra defines the reflection matrix up to the overall dressing phase only. This phase may be obtained by considering the crossing equation (see eg.\ \cite{HM}). Although it requires the dressing phase of the bulk S-matrix which at the moment is also not known. 

Finally, we show that the twisted quantum affine algebra and reflection matrix we have found in the rational $q\to1$ limit specializes exactly to the twisted Yangian of the $Y=0$ giant graviton and the reflection matrix associated to it, which for the reflection of arbitrary bound states was found by L. Palla \cite{LP}. Furthermore, we have casted the twisted Yangian of the $Y=0$ giant graviton in a very compact form \eqref{twistY} with the help of the outer automorphism $\B$ of the extended $\alg{su}(2|2)$ algebra. However, the explicit form \eqref{RLim} of the rational $q\to1$ limit of $\afB$ was somewhat surprising. Bearing on the Yangian limit of $\afQ$ found in \cite{BGM}, we were expecting to obtain the twisted secret charges constructed in \cite{VR}. However it turned out not to be the case, thus the role of the twisted secret charges of \cite{VR} remains unknown.   

Due to the high complexity of the bulk S-matrix we were unable to check the reflection equation in complete generality, however we did check a wide number of generic cases numerically and all of them were satisfied.

In this work we have used the reflection Hopf algebra formalism introduced in \cite{MR2}, however we have not stated explicitly the reflection automorphism of the algebra. We have only constructed the map between the representations of incoming and reflected states. This was sufficient for our purpose -- finding explicit form of the reflection matrix. Construction of the reflection automorphism requires a detailed analysis of the outer-automorphism group of $\afQ$ and thus is beyond of the scope of this paper. Nevertheless it is a very important question and deserves to be explored.

This work has revealed one more algebraic structure related to the deformed quantum affine algebra $\afQ$. The natural next step would be to explore boundary scattering for other boundary conditions, most notably the one of the same type as of the $Z=0$ giant graviton. It would also be interesting to apply the Bethe ansatz for these systems and derive their transfer matrices \cite{Murgan}. Furthermore, there has recently been a rapid development in the $q$-deformed Pohlmeyer reduced version of the $\ads$ superstring theory \cite{HT,HHM}, however the presence of boundaries in the Pohlmeyer reduced theories has not been much investigated. Thus it would be very interesting to see the effects of boundaries in such theories and find plausible links to the algebraic constructions considered in this work.


\paragraph{Acknowledgements.}

The authors would like to thank Niklas Beisert, Stefan Kolb, Alexander Molev, Rafael Nepomechie and Alessandro Torrielli 
for useful discussions. We especially thank Niall MacKay for many comments and suggestions on the manuscript.
T.M. would like to warmly thank the Max-Planck-Institut f\"ur Gravitationsphysik, 
Albert-Einstein-Institut in Potsdam where the main part of this work was performed. 
M. dL. thanks Swiss National Science Foundation for funding under the project number 200021-137616. 
V.R. also thanks the UK EPSRC for funding under grant EP/H000054/1.


\appendix

\section{Algebra maps} \label{appMap}

\paragraph{The $q\to1$ limit map.}

Algebra $\afQ$ in the conventional $q\to1$ limit specializes to the 
centrally extended $\mathfrak{sl}(2|2)$. Central elements $U$ and $V$ of the algebra 
in this limit become $U\to\mathcal{U}$ and $V\to1$. The representation labels 
$(a,b,c,d)$ \eqref{abcd} specialize to the usual non-deformed labels 
$(a,b,c,d)$ of \cite{BAnalytic}, while the affine ones become 
\begin{eqnarray}
\left(\tilde{a},\,\tilde{b},\,\tilde{c},\,\tilde{d}\right) & \to & \left(\alpha\tilde{\alpha}\, c,\,\alpha\tilde{\alpha}\, d,-\frac{a}{\alpha\tilde{\alpha}},-\frac{b}{\alpha\tilde{\alpha}}\right).
\end{eqnarray}
The natural choice of normalization for $\tilde{\alpha}$ is $1$. The explicit map between the generators of the algebras is
\begin{equation}
E_{i}\to\mathfrak{E}_{i},\quad F_{i}\to\mathfrak{F}_{i},\quad H_{i}\to\mathfrak{H}_{i}\qquad\mbox{for }i=1,2,3,
\end{equation}
and
\begin{equation}
E_{4}\to\alpha\,\mathfrak{F}_{321},\quad F_{4}\to-\alpha^{-1}\mathfrak{E}_{321},\quad H_{4}\to-\mathfrak{H}_{1}-\mathfrak{H}_{2}-\mathfrak{H}_{3}.
\end{equation}
The central charges are being mapped as
\begin{align}
C_1\to\mathfrak{C}, \qquad C_2\to\mathfrak{P}, \qquad C_3\to\mathfrak{K}.
\end{align}


\paragraph{The $\mathfrak{sl}(2|2)\to\mathfrak{su}(2|2)$ map.}

The map between these algebras reads as
\begin{eqnarray}
 &  & \mathfrak{E}_{1}\to\mathbb{L}_{2}^{1},\qquad\mathfrak{\mathfrak{F}}_{1}\to\mathbb{L}_{1}^{2},\qquad\mathfrak{H}_{1}\to-2\mathbb{L}_{1}^{1},\el
 &  & \mathfrak{E}_{2}\to\mathbb{Q}_{4}^{2},\qquad\mathfrak{\mathfrak{F}}_{2}\to\mathbb{G}_{2}^{4},\qquad\mathfrak{H}_{2}\to-\mathbb{L}_{1}^{1}-\mathbb{R}_{3}^{3}+\frac{1}{2}\mathbb{H},\el
 &  & \mathfrak{E}_{3}\to\mathbb{R}_{3}^{4},\qquad\mathfrak{F}_{3}\to\mathbb{R}_{4}^{3},\qquad\mathfrak{H}_{3}\to-2\mathbb{R}_{3}^{3}.
\end{eqnarray}
Their commutators are
\begin{eqnarray}
 &  & \mathfrak{E}_{32}\to\mathbb{Q}_{3}^{2},\qquad\mathfrak{E}_{21}\to\mathbb{Q}_{4}^{1},\qquad\mathfrak{E}_{321}\to\mathbb{Q}_{3}^{1},\el
 &  & \mathfrak{F}_{23}\to\mathbb{G}_{2}^{3},\qquad\mathfrak{F}_{12}\to\mathbb{G}_{1}^{4},\qquad\mathfrak{F}_{321}\to\mathbb{G}_{1}^{3}.
\end{eqnarray}
Finally, the central charges are being mapped as
\begin{align}
\mathfrak{C}\to2\mathbb{H}, \qquad \mathfrak{P}\to\mathbb{C}, \qquad \mathfrak{K}\to\mathbb{C}^\dagger.
\end{align}
For details on $\mathfrak{su}(2|2)$ see e.g.\ \cite{LeeuwThesis}.
 

\section{Symmetries of Y=0 giant graviton} \label{appYD3}

The $Y=0$ giant graviton preserves the $\mathfrak{su}(2|1)$ subalgebra
of the $\mathfrak{psu}(2|2)\ltimes\mathbb{R}^{3}$ and has no degrees
of freedom attached to the end of the spin chain \cite{HM}. 
The algebra $\h=\mathfrak{su}(2|1)$ is obtained from the $\g=\mathfrak{psu}(2|2)\ltimes\mathbb{R}^{3}$
by dropping the generators with bosonic indices $a,\, b,\, c,\;...=1$ (or equivalently with $a,\, b,\, c,\;...=2$). 
Thus the surviving (preserved) charges from a subgroup $\h=\{\mathbb{R}_{\alpha}^{\enskip\beta},\,\mathbb{L}_{1}^{\enskip1},\,\mathbb{Q}_{\alpha}^{\enskip2},\,\mathbb{G}_{2}^{\enskip\alpha},\,\mathbb{H}\}$,
while the broken charges form a subset $\m=\mathfrak{psu}(2|2)\ltimes\mathbb{R}^{3}\backslash \mathfrak{su}(2|1)$
consisting of $\{\mathbb{L}_{1}^{\enskip2},\;\mathbb{L}_{2}^{\enskip1},\;\mathbb{Q}_{\gamma}^{\enskip1},\;\mathbb{G}_{1}^{\enskip\gamma},\;\mathbb{C},\;\mathbb{C}^{\dagger}\}$. 

Since the Cartan-Killing form of the extended $\alg{su}(2|2)$ algebra is degenerate, 
the algebra does not have well-defined Casimir operator. 
Thus we cannot apply the formula \eqref{twist} naively\footnote{%
In the case of bulk scattering, one of the regularization is to use the $\alg{su}(2)$ outer automorphism \cite{BeisertYangian}.}.
In order to use \eqref{twist},  
the Casimir operator of $\mathfrak{su}(2|1)$ have to be enhanced by $\alg{u}(1)$ outer automorphism $\B$, 
\begin{align}
\mathbb{T}^{\alg{a}} = - {\R}_\gamma^\delta{\R}_\delta^\gamma + 2{\L}^1_1{\L}^1_1
 + {\Q}_\gamma^2{\G}_2^\gamma - {\G}_2^\gamma{\Q}_\gamma^2 - 2 \H\,\B\,. 
\end{align}
This extra charge is the Cartan generator of $\alg{su}(2)$ outer automorphism of the extended $\alg{su}(2|2)$ algebra
and it serves the hypercharge for generators,
\begin{align}
[\B,\Q_\alpha^a] & = + \tfrac{1}{2}\Q_\alpha^a,\qquad
[\B,\G_a^\alpha] = - \tfrac{1}{2}\G_a^\alpha,\qquad 
[\B,\C] = + \C,\qquad 
[\B,\C^\dagger] = -\C^\dagger,\nonumber\\
[\B,\L_a^b] & = [\B,\R_\alpha^\beta] = [\B,\H] = 0.
\end{align}
Then \eqref{twist} implies that the twisted Yangian charges governing the scattering of
the $Y=0$ giant graviton are 
\begin{align}
\widetilde{\mathbb{Q}}_{\alpha}^{1} & = \widehat{\mathbb{Q}}_{\alpha}^{1} - \tfrac{1}{4} \bigl[\mathbb{T}^\h,{\mathbb{Q}}_{\alpha}^{1}\bigr] ,
& \widetilde{\mathbb{L}}_{1}^{2} & = \widehat{\mathbb{L}}_{1}^{2} - \tfrac{1}{4} \bigl[\mathbb{T}^\h,{\mathbb{L}}_{1}^{2}] ,
& \widetilde{\mathbb{C}} & = \widehat{\mathbb{C}} - \tfrac{1}{4} \bigl[\mathbb{T}^\h,{\mathbb{C}}] ,\el
\widetilde{\mathbb{G}}_{1}^{\alpha} & = \widehat{\mathbb{G}}_{1}^{\alpha} - \tfrac{1}{4} \bigl[\mathbb{T}^\h,{\mathbb{G}}_{1}^{\alpha}] ,
& \widetilde{\mathbb{L}}_{2}^{1} & = \widehat{\mathbb{L}}_{2}^{1} - \tfrac{1}{4} \bigl[\mathbb{T}^\h,{\mathbb{L}}_{2}^{1}] ,
& \widetilde{\mathbb{C}}^{\dagger} & = \widehat{\mathbb{C}}^{\dagger} - \tfrac{1}{4} \bigl[\mathbb{T}^\h,{\mathbb{C}}^{\dagger}].\label{twistY}
\end{align}
The explicit forms of their co-products are 
\begin{eqnarray}
\Delta\widetilde{\mathbb{Q}}_{\alpha}^{1} & = & \widetilde{\mathbb{Q}}_{\alpha}^{1}\otimes1+1\otimes\widetilde{\mathbb{Q}}_{\alpha}^{1}+\mathbb{Q}_{\bar{\alpha}}^{1}\otimes\mathbb{R}_{\alpha}^{\bar{\alpha}}-\mathbb{L}_{2}^{1}\otimes\mathbb{Q}_{\alpha}^{2}+\mathbb{Q}_{\alpha}^{1}\otimes\K_{\alpha}^{\alpha}+\varepsilon_{\alpha\beta}\mathbb{C}\otimes\mathbb{G}_{2}^{\beta},\el
\Delta\widetilde{\mathbb{G}}_{1}^{\alpha} & = & \widetilde{\mathbb{G}}_{1}^{\alpha}\otimes1+1\otimes\widetilde{\mathbb{G}}_{1}^{\alpha}-\mathbb{G}_{1}^{\bar{\alpha}}\otimes\mathbb{R}_{\bar{\alpha}}^{\alpha}+\mathbb{L}_{1}^{2}\otimes\mathbb{G}_{2}^{\alpha}-\mathbb{G}_{1}^{\alpha}\otimes\K_{\alpha}^{\alpha}-\varepsilon^{\alpha\beta}\mathbb{C}^{\dagger}\otimes\mathbb{Q}_{\beta}^{2},\el
\Delta\widetilde{\mathbb{L}}_{1}^{2} & = & \widetilde{\mathbb{L}}_{1}^{2}\otimes1+1\otimes\widetilde{\mathbb{L}}_{1}^{2}-2\mathbb{L}_{1}^{2}\otimes\mathbb{L}_{1}^{1}-\mathbb{G}_{1}^{\gamma}\otimes\mathbb{Q}_{\gamma}^{2},\el
\Delta\widetilde{\mathbb{L}}_{2}^{1} & = & \widetilde{\mathbb{L}}_{2}^{1}\otimes1+1\otimes\widetilde{\mathbb{L}}_{2}^{1}+2\mathbb{L}_{2}^{1}\otimes\mathbb{L}_{1}^{1}-\mathbb{Q}_{\gamma}^{1}\otimes\mathbb{G}_{2}^{\gamma},\el
\Delta\widetilde{\mathbb{C}} & = & \widetilde{\mathbb{C}}\otimes1+1\otimes\widetilde{\mathbb{C}}+\mathbb{C}\otimes\mathbb{H},\el
\Delta\widetilde{\mathbb{C}}^{\dagger} & = & \widetilde{\mathbb{C}}^{\dagger}\otimes1+1\otimes\widetilde{\mathbb{C}}^{\dagger}-\mathbb{C}^{\dagger}\otimes\mathbb{H}.
\end{eqnarray}
Here $\K_{\alpha}^{\alpha}=\mathbb{R}_{\alpha}^{\alpha}+\mathbb{L}_{1}^{1}+\frac{1}{2}\mathbb{H}$
and $\bar{4}=3$, $\bar{3}=4$.


\global\long\def\nlin#1#2{\href{http://xxx.lanl.gov/abs/nlin/#2}{\tt nlin.#1/#2}}
\global\long\def\hepth#1{\href{http://xxx.lanl.gov/abs/hep-th/#1}{\tt hep-th/#1}}
\global\long\def\condmat#1{\href{http://xxx.lanl.gov/abs/cond-mat/#1}{\tt cond-mat/#1}}
\global\long\def\arXivid#1{\href{http://arxiv.org/abs/#1}{\tt arXiv:#1}}
\global\long\def\xmath#1{\href{http://arxiv.org/abs/math/#1}{\tt math/#1}}

\end{document}